\documentclass[usegraphicx]{mn2e}
\begin{document}
\title[Fanaroff-Riley dichotomy and the Malmquist bias]
{Fanaroff-Riley dichotomy of radio galaxies and the Malmquist bias}
\author[Singal \&  Rajpurohit]{Ashok K. Singal{$^1$} \& Kamlesh Rajpurohit{$^2$}\\
{$^1$}{Astronomy and Astrophysics Division, Physical Research Laboratory, 
Navrangpura, Ahmedabad - 380 009, India}\\
{$^2$}{Thuringer Landessternwarte (TLS), Sternwarte 5, 07778 Tautenburg, Germany}
\\E-mail: asingal@prl.res.in (AKS), kamlesh@tls-tautenburg.de (KR)}
\date{Accepted . Received ; in original form }
\maketitle
\begin{abstract}
We examine the possibility that a claimed dependence of  
the FR 1/2 break value in radio luminosity on the absolute magnitude of the optical host galaxy could be due to the Malmquist 
bias, where a redshift-luminosity correlation appears in a flux-limited sample because of an observational selection effect. 
In such a sample, the redshift dependence of a phenomenon could appear as a luminosity-dependent effect and may not be really 
representing an intrinsic property of the radio sample. We test this on the radio complete MRC (Molonglo Reference Catalog) 
sample, where Spearman rank correlation and Kendall rank correlation tests show that the correlations are indeed stronger 
between the redshift and the optical luminosity than that between the radio luminosity and the optical luminosity, suggesting that the 
latter correlation perhaps 
arises because of the Malmquist Bias. We further show that similar effects of the Malmquist bias could also be 
present elsewhere in other correlations claimed in the literature between the radio luminosity and other observed properties of 
FR1 and FR2 sources.
\end{abstract}
\section{Introduction}
One of the robust correlations in extragalactic radio astronomy is between the morphology types of 
radio galaxies and their radio luminosity, first noted by Fanaroff \& Riley (1974). 
In their archetypal paper Fanaroff \& Riley pointed out that almost all radio galaxies below 
$P_{178} = 2 \times 10^{25}$ W Hz$^{-1}$ sr$^{-1}$ (for Hubble constant 
$H_{0}=50\,$km~s$^{-1}$\,Mpc$^{-1}$), are edge-darkened (called type 1) in their brightness distribution, 
while all radio galaxies above this luminosity limit are more or less edge-brightened (called type 2). 
This correlation has withstood the test of time (Miley 1980; Antonucci 1993; Urry \& Padovani 1995; 
Saripalli 2012), though the subsequent observations have shown that the transition is not really so sharp,   
encompassing a band $\sim 2$ order of magnitude wide in the radio luminosity (see e.g., Baum et al. 1995 and the references 
therein). Additionally, there is also a claim (Owen \& Ledlow 1994; Ledlow \& Owen 1996) that the FR1/2  break value in the radio 
luminosity may itself be a function of the optical luminosity of the host galaxy.  

However, because of a strong redshift-luminosity correlation (Malmquist bias) 
in a flux-limited sample, any effect related with redshift can appear as a luminosity-dependent effect. 
A question could then arise -- is the construed dependence of the break value in the radio 
luminosity on the optical luminosity likely to be due to the Malmquist bias? Moreover, could some of the other 
correlations claimed in the literature 
between the radio luminosities and other observed properties of FR1 and FR2 sources, e.g. the line or continuum 
optical luminosities of the host galaxies etc., be partially or even wholly results of the Malmquist bias?  
We investigate these possibilities here. 
\section{The samples and the data}
To study the effect of the Malmquist bias on the shift observed in FR1/2 break value in the radio luminosity with a change in 
the optical luminosity of the host galaxy, we have chosen the essentially complete MRC  
sample (Kapahi et al. 1998) with $S_{408} \ge 0.95$ Jy, and which has the required radio and optical information. 
The total sample comprises 550 sources, with 111 of them being quasars and the remainder radio galaxies. 
To quantitatively distinguish between  FR1 and 2, following Fanaroff \& Riley (1974), we classify a radio galaxy 
as FR1 if the separation between the points of peak intensity 
in the two lobes is smaller than half the largest size of the source. Similarly FR2 is the one in which the 
separation between the points of peak 
intensity in the two lobes is greater than half the largest size of the source. 
This is equivalent to having the ``hot spots'' nearer to (FR1) or further away from (FR2) 
the central optical galaxy than the regions of diffuse radio emission.
We have examined the radio maps and using the above criteria have classified each source  
into either of the two types (FR1 and 2). There were a small number of sources that had a doubtful classification, 
we have dropped them from our analysis and that should not be too detrimental to our conclusions.
Optical identifications for the sample are complete up to a red magnitude of $\sim 24$ 
or a {\em K} magnitude of  $\sim 19$. The still non-identified cases are expected to be at high redshifts 
($z \stackrel{>}{_{\sim}}1$), and therefore of high radio luminosities as well, 
it is not likely that many FR1 types will be amongst those. 
The other sample we have examined for the Malmquist bias is that of Zirbel \& Baum (1995) 
where they have compiled data on FR1s and FR2s from literature 
and showed a correlation between the radio and optical luminosities of sources in their sample.
\begin{figure}
\scalebox{0.28}{\includegraphics{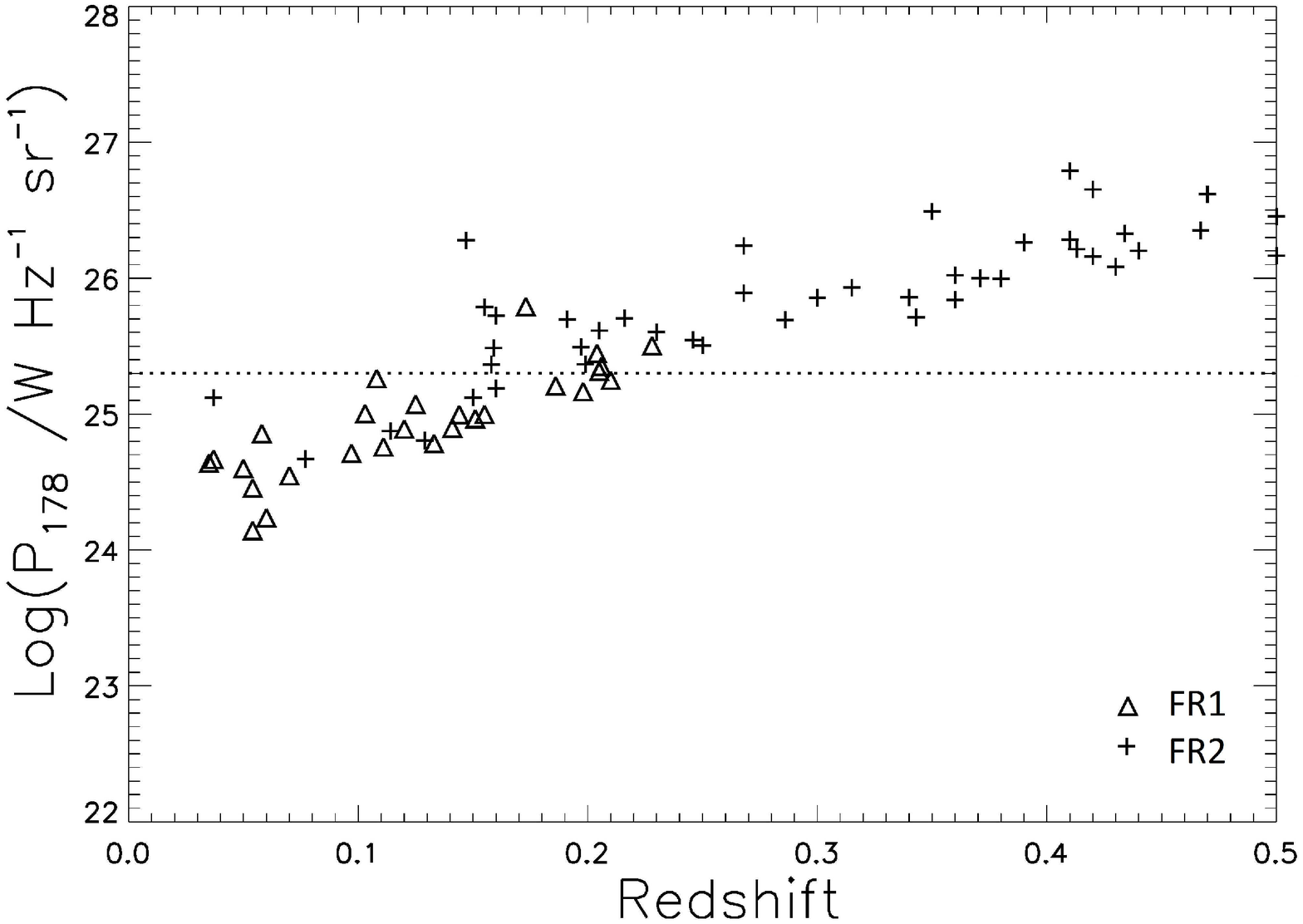}}
\caption{A plot of radio luminosity against redshift, showing a correlation due to the Malmquist bias, 
for both FR1 and 2 type sources in the MRC sample. The dotted line indicates the FR1/2 break-luminosity 
value from Fanaroff \& Riley (1974).}
\end{figure}
\section{Results and discussion}
Most or perhaps all FR1/2 radio data in the literature are from flux-limited samples, in which Malmquist bias will undoubtedly be 
present, showing a strong redshift-radio luminosity correlation in the data. Fig. 1 shows such a plot for our MRC data. 
A similar bias in the sample could as well be there even in the plots of optical absolute magnitudes of host galaxies 
against redshifts. In most of the  samples considered one does not have 100\% identifications with spectroscopic redshift 
measurements and often there is an upper limit on the apparent magnitude for optical identifications of galaxies. 
In such a case Malmquist bias would be present in the optical data as well. Perhaps one way to resolve it would be if we have 
samples where 100\% identifications with spectroscopic redshifts are available. But in the MRC sample incomplete 
optical identifications does not seem to be problem because 
more or less all FR1s have been optically identified. Of course there could also be some genuine correlation of 
optical luminosity with redshift, e.g., if there were a cosmic evolution of the optical luminosity of the host galaxy.
In such cases any plot of the radio luminosities vs. absolute visual magnitudes of the galaxies will, through their common denominator
of redshift, display a correlation. This correlation could be very strong,  even though the optical luminosity might not necessarily be 
having any intrinsic correlation with the radio luminosity of the source. It is difficult to
untangle a Malmquist bias from an intrinsic correlation. Any attempt to eliminate effects of a redshift correlation 
could inadvertently remove also some genuine effects of radio luminosity. Only if there is sufficient data in $P-z$ 
plane so that one could vary parameters $P$ and $z$ independently, could one disentangle the two effects.

\begin{figure}
\scalebox{0.28}{\includegraphics{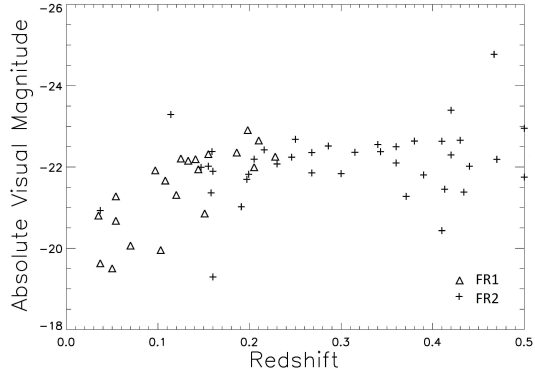}}
\caption{A plot of the absolute visual magnitude of the host galaxy as a function of redshift for both FR1 
and 2 type sources in the MRC sample.}
\end{figure}

Fig. 2 shows the presence of a correlation of the absolute visual magnitudes of the host galaxy with redshift, 
especially for FR1 types, in the MRC data. Irrespective of the ultimate cause, there is no denying that a discernible 
correlation exists between absolute visual magnitude and redshift, more so among FR1s. In Fig. 3 
we see how the correlations for the MRC data seen in Figs. 1 and 2 show up as semblance of a correlation between radio 
luminosities and the absolute visual magnitudes of the host galaxies, especially for FR1 types. 
\begin{figure}
\scalebox{0.28}{\includegraphics{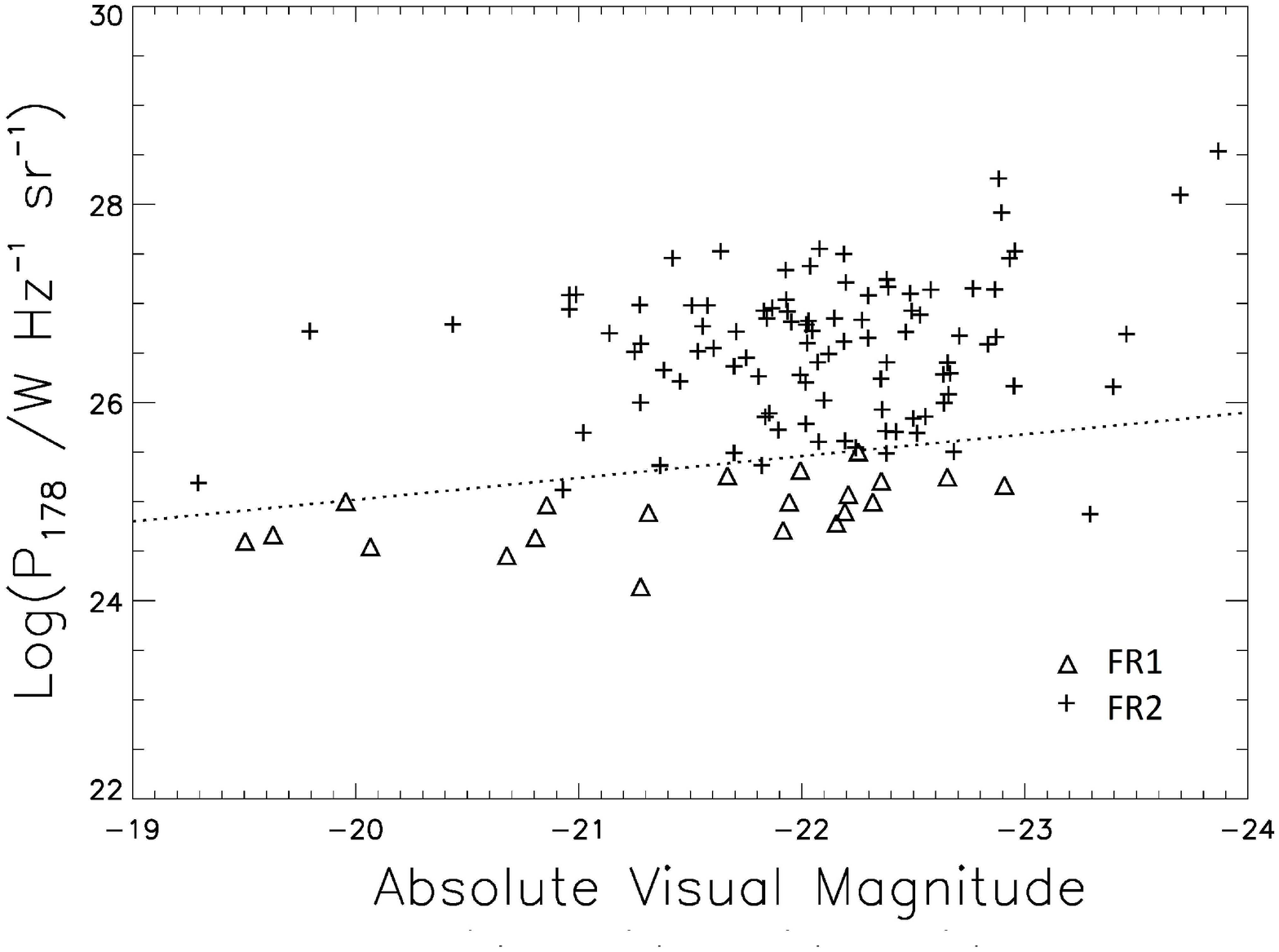}}
\caption{A plot of the radio luminosity against absolute visual magnitude of the host galaxy 
for both FR1 and 2 type sources in the MRC sample. The dotted line indicates the FR1/2 break-radio luminosity value.}
\end{figure}
Among FR2s the $M-z$ correlation, if any, is much weaker (Fig. 2), the same of course is reflected as a weaker  
$P-M$ correlation among FR2s in Fig.~3.

\begin{table*}
\caption{Spearman rank correlation $\rho$ and Kendall rank correlation $\tau$ between various pairs.}
\begin{tabular}{@{}ccccccccccccccccc}
\hline
Sample & FR type &&&$\rho(P-z)$ & $\rho(M-z)$ & $\rho(P-M)$  & $\sigma_{\rho}$ &&& $\tau(P-z)$ & 
$\tau(M-z)$ & $\tau(P-M)$ & $\sigma_{\tau}$\\
(1)&(2)&&&(3)&(4)&(5)&(6)&&&(7)&(8)&(9)&(10)\\
\hline
MRC & FR1 &&& 0.82  & 0.83  & 0.67 & 0.14 &&& 0.65& 0.63 & 0.49& 0.16\\
MRC & FR2 &&& 0.92  & 0.22  & 0.16 & 0.06 &&& 0.77& 0.15 & 0.11& 0.07\\\\
ZS &  FR1 &&& 0.85  & 0.75  & 0.76 & 0.08 &&& 0.69& 0.57 & 0.57& 0.09\\
ZS &  FR2 &&& 0.94  & 0.87  & 0.89 & 0.06 &&& 0.80& 0.67 & 0.69& 0.06\\
\hline
\end{tabular}
\end{table*}

Table 1 shows a Spearman rank correlation $\rho$ and Kendall rank correlation $\tau$ between pairs of various parameters, 
organized in the following manner. 
(1) Sample used.
(2) FR type of the sub-sample.
(3) Spearman rank correlation in radio luminosity and redshift.
(4) Spearman rank correlation in optical absolute magnitude and redshift.
(5) Spearman rank correlation in radio luminosity and optical absolute magnitude.
(6) Standard deviation in Spearman rank correlations.
(7) Kendall rank correlation in radio luminosity and redshift.
(8) Kendall rank correlation in optical absolute magnitude and redshift.
(9) Kendall rank correlation in radio luminosity and optical absolute magnitude.
(10) Standard deviation in  Kendall rank correlations.
Standard deviation in each case is calculated under the null hypothesis where the rank correlation between the respective 
pair of parameters is zero with the correlation coefficient having a normal distribution about the zero mean (Babu \& Feigelson 1996).   

We notice two things from Table 1 for the MRC sample. The $M-z$ and $P-M$ correlations are much stronger as well as 
statistically  more significant for FR1s as compared to FR2s, as seen from comparatively much higher values of $\rho$ 
and $\tau$ for FR1s. It was already apparent from Figs. 2 and 3 qualitatively, but now we have a more quantitative 
measure of the correlations and their statistical significance. But what was not that much apparent from Figs. 2 and 3 
was that the $M-z$ correlations in Table 1 are stronger and statistically more significant than the 
$P-M$ correlations, both for FR1s and FR2s. Therefore it is more likely that the primary correlation is between 
optical luminosity and redshift and that the weaker correlation between radio and optical luminosities arises as a 
consequence of the Malmquist Bias. 

If the FR1/2 break-radio luminosity had a sharp value (as stated by Fanaroff \& Riley 1974) then Fig. 3 would contain only a horizontal 
line with no apparent $P-M$ correlation. However a spread in the break-radio luminosity value will appear as 
a corresponding spread in the $z$ value due to the Malmquist bias. This in turn would show up as a spread in optical luminosity 
for a sample having a correlation in absolute magnitudes and redshift. Thereby the band of FR1/2 break in the radio luminosity would 
get stretched into a correlation between the radio luminosity of the source and the optical luminosity of its host galaxy. 
In case of a strong optical 
luminosity-redshift relation, the correlation coefficient could be almost as strong as for the radio luminosity-redshift case, 
since all luminosities have mathematically the same relation with redshift. A suspicion of such an artificial 
correlation could be there in the claimed correlation by Owen \& Ledlow (1994) and Ledlow \& Owen (1996), where the slope of the 
observed correlation is such that a change of optical magnitude of $\sim5$ in absolute magnitude corresponds to a change of radio 
luminosity of a factor of about $\sim 10^2$. Such a thing could be expected because of the Malmquist bias, 
since both radio and optical luminosities would change by the same factor as a function of redshift. After all it might not be a 
mere coincident that the correlation between radio and optical luminosities is almost exactly the same as expected from their 
redshift dependence. 

Also in the data of Owen \& Ledlow (1994) and Ledlow \& Owen (1996), the upper luminosity envelope for FR2 case too 
seems to be related to the absolute optical magnitude by 
an almost exactly similar slope. Therefore we think that the claimed correlation does not represent an actual intrinsic 
correlation but could merely be result of the Malmquist bias through their redshift dependence. Then any further theoretical 
interpretation would also need to be accordingly revised or at least looked at afresh.

Similar effects of the Malmquist bias could also be seen in some other correlations claimed between the radio luminosity and 
some other parameters like the optical line luminosity (Zirbel \& Baum 1995, figure 9). While the authors have mentioned 
the effects of redshift on radio and optical luminosities, but they have not really tried to remove them. 
\begin{figure}
\scalebox{0.35}{\includegraphics{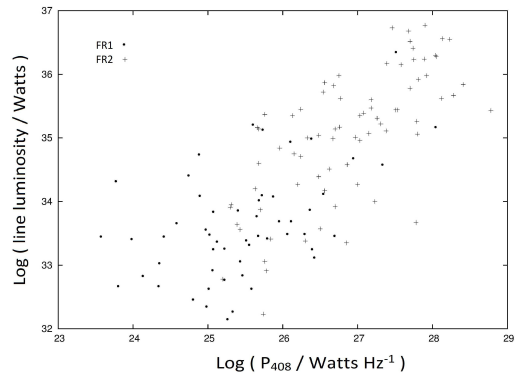}}
\caption{A plot of the line luminosity of the host galaxy against radio luminosity at 408 MHz 
for both FR1 and 2 type sources in the Zirbel \& Baum (1995) sample.}
\end{figure}

\begin{figure}
\scalebox{0.4}{\includegraphics{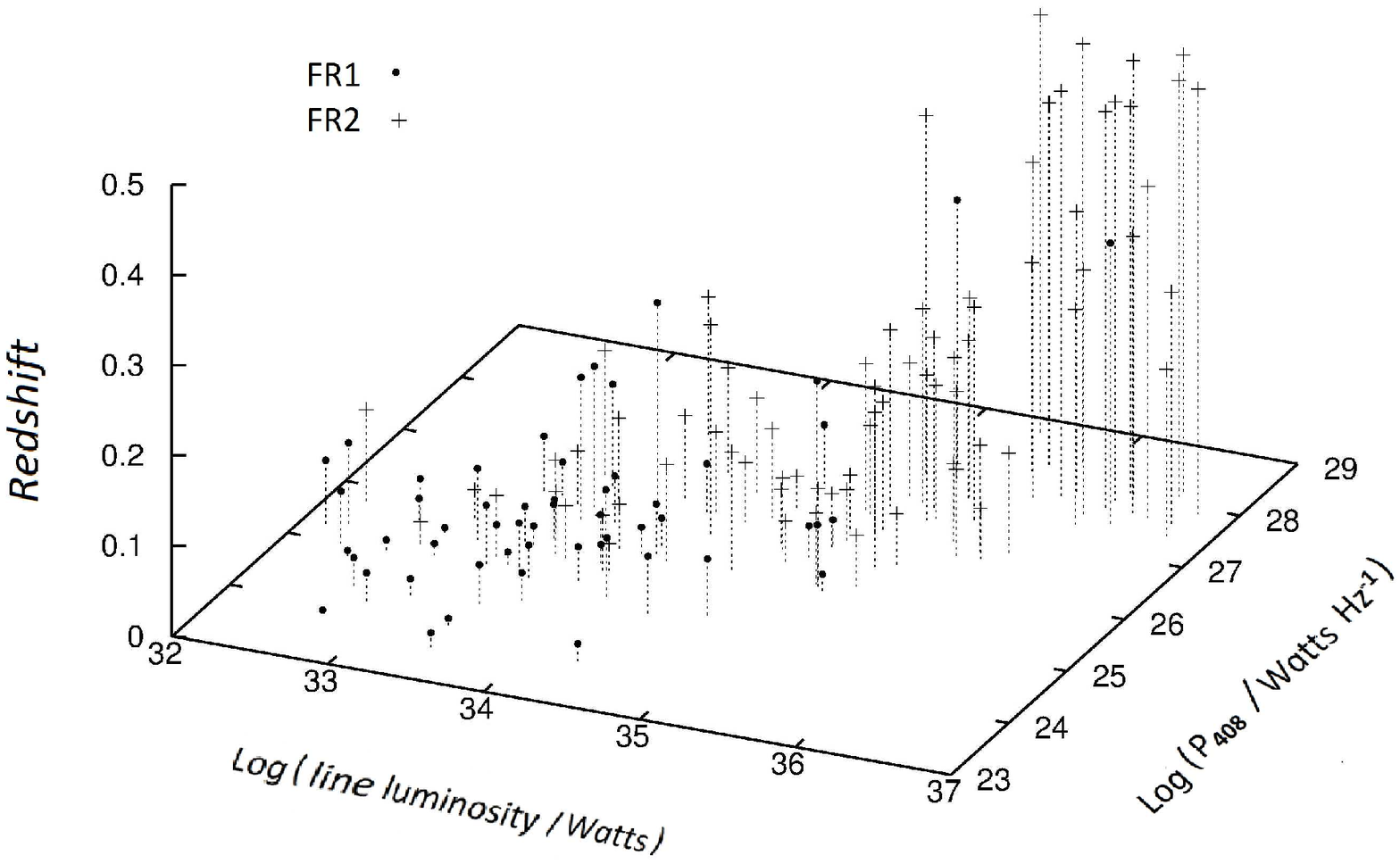}}
\caption{A 3d-plot of the redshift in the line luminosity-radio luminosity plane corresponding to Fig. 4,  
for both FR1 and 2 type sources in the Zirbel \& Baum (1995) sample.}
\end{figure}
\begin{figure}
\scalebox{0.38}{\includegraphics{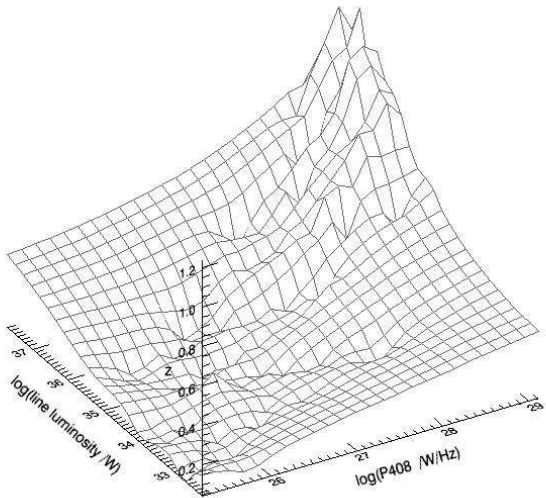}}
\caption{A 3d surface-plot of the redshift distribution in the line luminosity-radio luminosity plane corresponding to Fig. 4,  
for FR2 type sources.}
\end{figure}

In  Fig. 4 here we show a plot of the line luminosity of the host galaxy against radio 
luminosity at 408 MHz for the sample of FR1s and FR2s from Zirbel \& Baum (1995, table 11). The correlation seems 
to be quite tight between the two quantities. However this all could merely be a manifestation of the Malmquist bias 
which can be seen from Figs. 5 and 6 where we have plotted distribution of redshift ($z$), 
in 3d-plots in the radio luminosity -- optical line luminosity plane, for the same sources as in Fig. 4. 
While in Fig. 5 we have restricted the redshift range to $z<0.5$ to avoid too much 
compression of the scale for FR1s, Fig. 6 shows a 3-d surface plot for FR2s covering their full redshift range. 
From these two figures it seems that the apparent correlation seen in Fig. 4 
was perhaps because the higher radio luminosity sources due to the Malmquist bias are predominantly picked at 
higher redshifts and which necessarily will also be of higher optical line luminosities so as to be able to find their place in the 
sample at those high redshifts. Then the perceived correlation in Fig. 4 would have arisen mainly due to this selection effect and 
might not represent any real intrinsic correlation in the source properties. From Table 1 also we see that both $M-z$ and $P-M$ 
correlations are approximately of equal strength and of comparable statistical significance. Considering that, one cannot 
unambiguously say that it is an intrinsic correlation between radio and optical luminosities, as generally stated in the 
literature. Instead it could as well be a 
correlation between optical luminosity and redshift, masquerading as a $P-M$ correlation due to the Malmquist bias. The question 
could not decisively be settled based on any amount of arguments, sans actual observational data,  
and it could be a non-trivial task to isolate true 
correlation, if any, between the two luminosities in the presence of a Malmquist bias. 

\begin{figure}
\scalebox{0.58}{\includegraphics{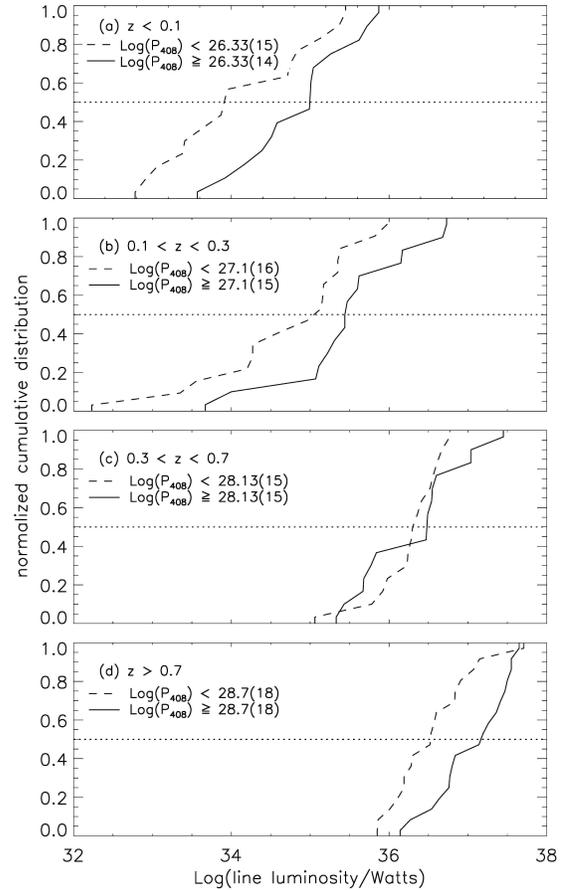}}
\caption{Correlation of the optical luminosity with radio luminosity of FR2s investigated from their normalized 
cumulative distributions for different radio luminosity bins within various redshift zones. The dotted lines intersect the 
cumulative plots at their median values. The number of sources in each radio luminosity bin are given within parentheses.}
\end{figure}

\begin{figure}
\scalebox{0.58}{\includegraphics{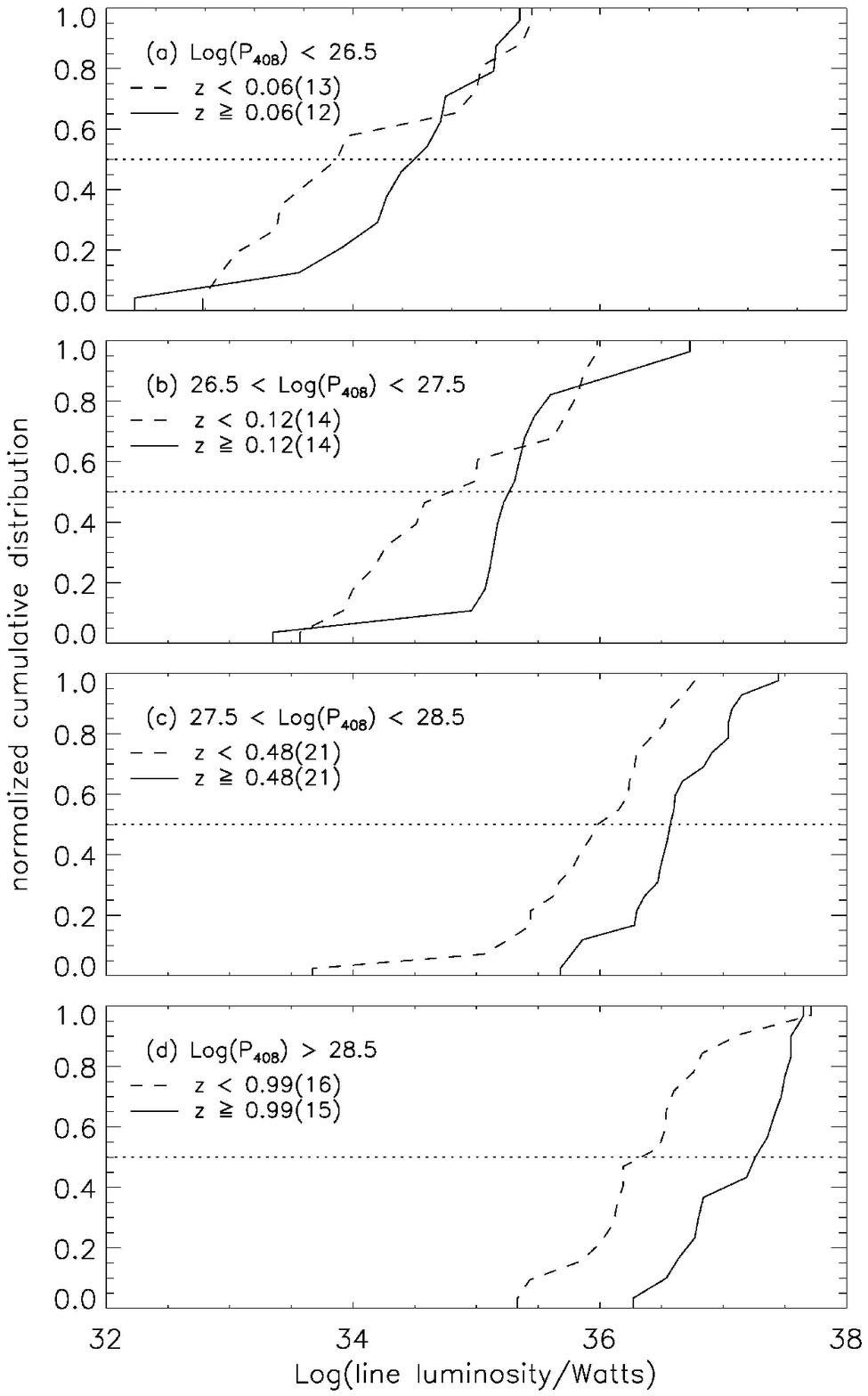}}
\caption{Correlation of the optical luminosity with redshift  of FR2s investigated from their normalized cumulative distributions
for different redshift bins within various radio luminosity zones. The dotted lines intersect the cumulative plots at their median 
values. The number of sources in each reshift bin are given within parentheses.}
\end{figure}

How to disentangle the two effects? As an illustration we quote from the literature a case where the Malmquist bias was 
successfully removed in the dependence of linear sizes on redshift and radio luminosity of radio galaxies and steep spectrum quasars. 
An earlier claim (Kapahi 1987) was that both radio galaxies and quasars have a similar radio size dependence on redshift. 
These results were marred by the Malmquist bias since size dependence, if any, on the radio luminosity  was not 
considered. However, using a much bigger sample, (Singal 1993) 
could separate the size dependences on luminosity and redshift by varying them independently and showed that the 
radio galaxies and quasars differed very much from each other in their size dependences on redshift and radio 
luminosity. Such was not possible to discern earlier due to a Malmquist bias.
\begin{figure}
\scalebox{0.45}{\includegraphics{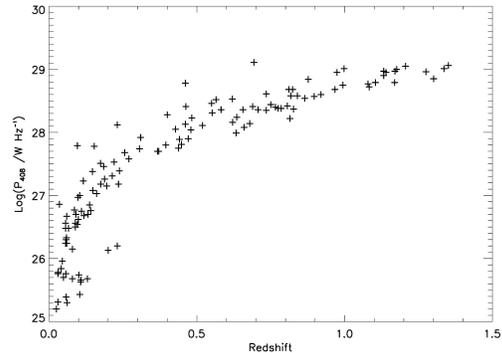}}
\caption{Correlation of the radio luminosity with redshift of FR2s from Zirbel \& Baum (1995).}
\end{figure}

We attempt here a similar separation by first intersecting the 3-d surface in Fig. 6 by three horizontal 
planes at $z=0.1$, 0.3 and 0.7, thereby dividing it into 4 redshift regions, with roughly similar number of sources in each region. 
Then within each redshift region we plotted the normalized cumulative distribution of the optical line luminosity 
for two radio luminosity bins. Due to the lack of sufficient sources, the data was   
divided in only two bins, choosing the median value of radio luminosity in that redshift range as the dividing line between the two bins. 
Fig. 7 shows the plots. Though the cumulative plots may have certain overlap at some places, but without an exception, 
the median value of the optical line luminosity is always higher for the higher radio luminosity bin. Thus we definitely see 
a direct correlation between the optical line luminosity and the radio luminosity within each redshift range and which might appear to 
support the claims of Zirbel \& Baum (1995). However a similar division into four radio luminosity regions and plots of 
the normalized cumulative distribution of the optical line luminosity for different redshift bins (Fig. 8) shows 
a similar correlation between the optical line luminosity and redshift, with 
the median value of the optical line luminosity being always higher for the higher redshift bin. 
It is clear from Figs. 7 and 8 that from the present data alone one cannot unambiguously decide whether the optical line luminosity 
is intrinsically correlated with the radio luminosity, redshift or both of them and how much might be the effect of the Malmquist bias. 
The Malmquist bias is too strong in the present data, due to a tight correlation between redshift and radio luminosity (Fig. 9), to 
disentangle the two. 
One needs much more data at different radio flux-density levels, so that one could get wide range of radio luminosities within 
narrow redshift ranges (and vice versa) to successfully disentangle the two effects.

\end{document}